\documentclass[sigconf]{acmart}

\usepackage[utf8]{inputenc}
\usepackage{listings}
\usepackage{enumitem}
\usepackage{xcolor}
\usepackage{amsmath}
\usepackage{cleveref}
\usepackage{subcaption}
\usepackage{algorithm}
\usepackage{algpseudocode}
\usepackage{pifont}
\usepackage{listings, listings-rust}

\algnewcommand\algorithmicforeach{\textbf{for each}}
\algdef{S}[FOR]{ForEach}[1]{\algorithmicforeach\ #1\ \algorithmicdo}\algrenewcommand\algorithmicrequire{\textbf{Input:}}
\algrenewcommand\algorithmicensure{\textbf{Output:}}

\crefformat{section}{\S#2#1#3}
\crefname{figure}{Figure}{Figures}
\crefname{table}{Table}{Tables}
\crefname{algorithm}{Algorithm}{Algorithms}

\newcommand{\code}[1]{\texttt{{\small \detokenize{#1}}}}

\newcommand{\data}[1]{\textcolor{black}{#1}}

\newcommand{\loancold}[1]{\textcolor{blue}{chenghao:#1}}

\copyrightyear{2024}
\acmYear{2024}
\setcopyright{acmlicensed}\acmConference[ICSE '24]{2024 IEEE/ACM 46th International Conference on Software Engineering}{April 14--20, 2024}{Lisbon, Portugal}
\acmBooktitle{2024 IEEE/ACM 46th International Conference on Software Engineering (ICSE '24), April 14--20, 2024, Lisbon, Portugal}
\acmPrice{15.00}
\acmDOI{10.1145/3597503.3623352}
\acmISBN{979-8-4007-0217-4/24/04}

\AtBeginDocument{%
  \providecommand\BibTeX{{%
    \normalfont B\kern-0.5em{\scshape i\kern-0.25em b}\kern-0.8em\TeX}}}

\title{Demystifying Compiler Unstable Feature Usage and Impacts in the Rust Ecosystem}

\date{}

\author{Chenghao Li}
\orcid{0009-0009-2920-0552}
\affiliation{%
\institution{Zhejiang University}
\city{Hangzhou}
\country{China}
}
\email{loancold@zju.edu.cn}

\author{Yifei Wu}
\orcid{0009-0002-2783-8623}
\affiliation{
\institution{Zhejiang University}
\city{Hangzhou}
\country{China}
}
\email{3190106075@zju.edu.cn}

\author{Wenbo Shen}
\orcid{0000-0003-2899-6121}
\authornote{Corresponding Author.}
\affiliation{
\institution{Zhejiang University}
\city{Hangzhou}
\country{China}
}
\email{shenwenbo@zju.edu.cn}

\author{Zichen Zhao}
\orcid{0009-0007-1222-2371}
\affiliation{
\institution{Zhejiang University}
\city{Hangzhou}
\country{China}
}
\email{zhaozichen@zju.edu.cn}

\author{Rui Chang}
\orcid{0000-0002-0178-0171}
\affiliation{
\institution{Zhejiang University}
\city{Hangzhou}
\country{China}
}
\email{crix1021@zju.edu.cn}

\author{Chengwei Liu}
\orcid{0000-0003-1175-2753}
\affiliation{
\institution{Nanyang Technological University}
\city{Singapore}
\country{Singapore}
}
\email{chengwei001@e.ntu.edu.sg}

\author{Yang Liu}
\orcid{0000-0001-7300-9215}
\affiliation{
\institution{Nanyang Technological University}
\city{Singapore}
\country{Singapore}
}
\email{yangliu@ntu.edu.sg}

\author{Kui Ren}
\orcid{0000-0002-1969-2591}
\affiliation{
\institution{Zhejiang University}
\city{Hangzhou}
\country{China}
}
\email{kuiren@zju.edu.cn}

\begin{document}

\begin{abstract}
Rust programming language is gaining popularity rapidly in building reliable and secure systems due to its security guarantees and outstanding performance.
%
To provide extra functionalities, the Rust compiler introduces Rust unstable features (RUF) to extend compiler functionality, syntax, and standard library support.
However, these features are unstable and may get removed, introducing compilation failures to dependent packages.
Even worse, their impacts propagate through transitive dependencies, causing large-scale failures in the whole ecosystem.
Although RUF is widely used in Rust, previous research has primarily concentrated on Rust code safety, with the usage and impacts of RUF from the Rust compiler remaining unexplored.
Therefore, we aim to bridge this gap by systematically analyzing the RUF usage and impacts in the Rust ecosystem.
%
We propose novel techniques for extracting RUF precisely, and to assess its impact on the entire ecosystem quantitatively, we accurately resolve package dependencies.
We have analyzed the whole Rust ecosystem with \data{590K} package versions and \data{140M} transitive dependencies. 
Our study shows that the Rust ecosystem uses \data{1000} different RUF, and at most \data{44\%} of package versions are affected by RUF, causing compiling failures for at most \data{12\%} of package versions.
To mitigate wide RUF impacts, we further design and implement a RUF-compilation-failure recovery tool that can recover up to \data{90\%} of the failure.
We believe our techniques, findings, and tools can help stabilize the Rust compiler, ultimately enhancing the security and reliability of the Rust ecosystem.


%

\end{abstract}



\maketitle

\section{Introduction}

\begin{figure*}[!t]
\centering
\includegraphics[width=0.9\textwidth]{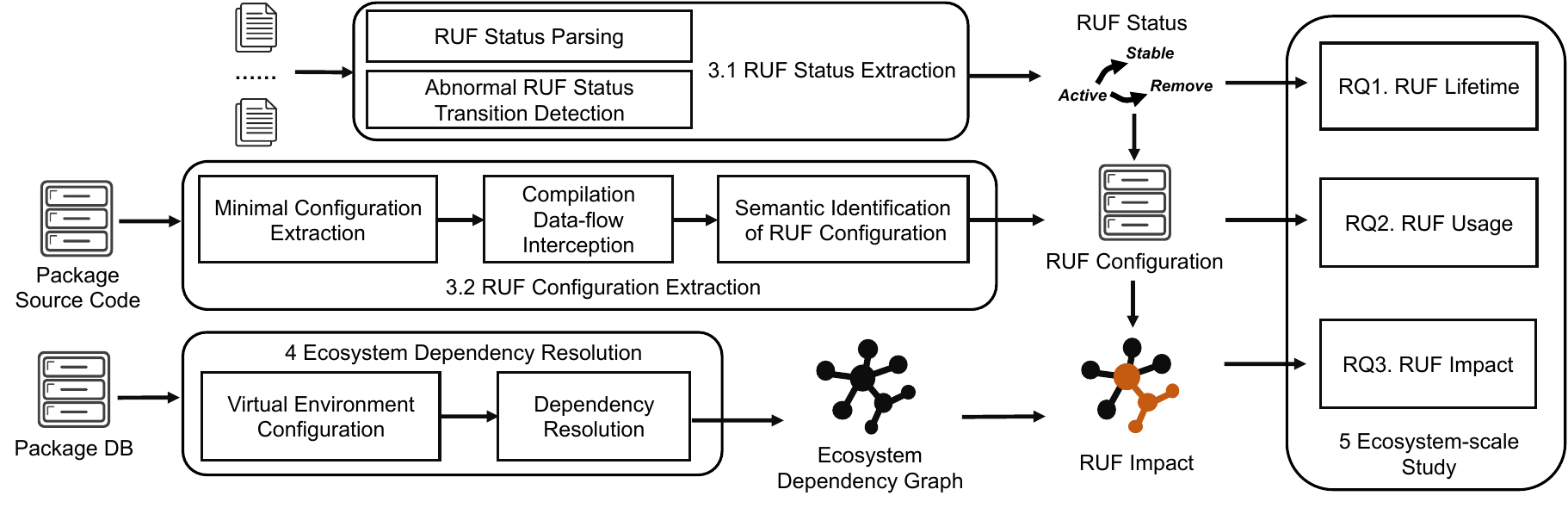}
\caption{Architecture overview of our work.}
\label{fig:impl_arch}
\end{figure*}

In recent years, Rust~\cite{rust} has been widely used to build reliable software productively, with its unique design for security and performance. 
Rust enforces memory safety and type safety via the ownership mechanism without garbage collection~\cite{rust_kernel, rust_case_study}. 
This makes Rust a great option for developing secure and efficient applications and frameworks, such as browsers, virtualization, database, and game software stack \cite{rocket, stdweb, rust_wasm, servo_paper, firecracker, redox, TiKV, Bevy}.
Large projects, including Android and Linux, also integrate Rust into their main projects for security concerns.
While the software ecosystem brings convenience to software development, it can also introduce potential reliability and security concerns to the software \cite{OSS_Review, JAVAOSS}.

Although Rust is being increasingly adopted, the Rust ecosystem is still young, and many problems within the ecosystem have not been well studied.
Previous studies on Rust security primarily focus on security threats caused by developers breaking Rust compiler security checks \cite{rust_used_safely, rust_rudra, rust_galeed, rust_beyond_safety, rust_belt,rust_crust, rust_understanding_safety}, but ignores the problems of the compiler itself.
We observe that the compiler allows developers to use Rust unstable features (RUF) to extend the functionalities of the compiler. 
However, RUF may introduce vulnerabilities to Rust packages~\cite{cve:error_type_id}, and removed RUF will make packages using it suffer from compilation failure. 
Even worse, the compilation failure can propagate through package dependencies, causing potential threats to the entire ecosystem.
Although RUF are widely used by Rust developers, unfortunately, to the best of our knowledge, its usage and impacts on the whole Rust ecosystem have not been studied so far.

To fill this gap, this paper conducts the first in-depth study to analyze RUF usage and its impacts on the whole Rust ecosystem.
We begin by extracting RUF definitions from the compiler and their usage from packages. Following that, we resolve all dependencies across the entire ecosystem, allowing us to quantify RUF impacts on an ecosystem scale.
Though conceptually simple, we must resolve three challenges to achieve the analysis.

First, it is hard to extract RUF. There is no official documentation to specify RUF definitions and usage, and they frequently change with compiler release updates. As a result, the syntax of RUF is not unified, and no existing technique can extract RUF accurately. To resolve this challenge, we develop new techniques to track all RUF supported by each version of the Rust compiler and all RUF used by Rust packages in the ecosystem. 
Second, the Rust package manager cannot be used to analyze the entire ecosystem in an acceptable time. Existing dependency-resolving techniques \cite{npm_vul_impact, small_world, npm_vul} fail to cover all dependency types and use approximate resolution algorithms, leading to inaccurate dependency resolution. Therefore, we propose an accurate Ecosystem Dependency Graph (EDG) generator to resolve dependencies in the Rust ecosystem.
Third, it is challenging to quantify RUF impacts precisely. RUF impacts other packages conditionally, determined by both RUF configuration and dependency attributes. 
Therefore, we propose the semantic identification of RUF configuration to precisely identify RUF usage and convey its impacts.

By conquering the above challenges, we analyze all packages on the official package database \code{crates.io} and resolve \data{592,183} package versions to get \data{139,525,225} transitive dependencies and \data{182,026} RUF configurations.
Our highlighted findings are:
1) About half of RUF (47\%) are not stabilized in the latest version of the Rust compiler;
2) \data{72,132 (12\%)} package versions in the Rust ecosystem are using RUF, and \data{90\%} of package versions among them are still using unstabilized RUF;
3) Through dependency propagation, RUF can impact at most \data{259,540 (44\%)} package versions, causing at most \data{70,913 (12\%)} versions to suffer from compilation failure. 
To further mitigate RUF impacts, we propose a new technique to detect RUF impacts on packages and recover them if RUF make them suffer from compilation failure. Theoretically, \data{90\%} compilation failure caused by RUF impacts can be recovered.
With our novel techniques, our work further advances the state of the art in analyzing RUF usage and impacts. 
Our unique findings reveal the stability problems of the Rust compiler, and based on these insights, we offer practical mitigation tools and suggestions to foster a more reliable Rust ecosystem.

The contributions of this paper can be summarized as follows:
\begin{itemize}[noitemsep, leftmargin=*]
    \item \textbf{New Study.} We are the first to investigate the usage and impacts of RUF in the Rust ecosystem. Our study demystifies RUF from three aspects: How RUF evolve with compiler upgrades, how Rust packages use RUF, and how RUF impact packages through dependencies. 
    
    \vspace{+2pt}
    \item \textbf{New Technique.} We propose novel techniques to extract RUF and determine RUF impacts in the Rust ecosystem.
    First, we propose the \textit{compilation data-flow interception} to extract RUF usage of Rust packages precisely. 
    Second, we propose \textit{semantic identification of RUF configuration} to recognize the semantics among RUF configurations.
    Third, we propose \textit{ecosystem dependency graph generator} to efficiently and accurately determine RUF impacts through dependencies.
    Last, we propose \textit{RUF impact mitigation} to help recover from compilation failure.
    
    \vspace{+2pt}
    \item \textbf{Ecosystem-scale Analysis.}
    We conduct the first ecosystem-scale analysis with \data{140} million transitive dependencies counted to reveal the RUF impacts. In our analysis, we find that RUF can impact almost half (\data{44\%}) of the Rust ecosystem, revealing the significant impacts and corresponding problems of RUF in the Rust ecosystem.
    
    \vspace{+2pt}
    \item \textbf{Community Contributions.}
    We design and implement a RUF-compilation-failure recovery tool that can recover up to 90\% of RUF compilation failures.
    We open source all RUF analysis implementations, tools, and data sets to the public to help the community track and fix RUF problems \footnote{\url{https://doi.org/10.5281/zenodo.8289375}}.
    
\end{itemize}

The core architecture of the paper is shown in~\cref{fig:impl_arch}.
We introduce background knowledge in~\cref{sec:background}.
In~\cref{RUF Extraction}, we propose techniques to extract RUF status and usage.
We then propose techniques to resolve dependencies and quantify the RUF impacts in~\cref{RUF Impact Resolution}.
In~\cref{Ecosystem-scale Study}, we further conduct an ecosystem-scale study on RUF and answer three research questions.
We continue by discussing our RUF impact mitigation techniques and suggestions in~\cref{sec:Mitigation}.
Related work is presented in~\cref{sec:related}.
Lastly, we conclude the whole paper in~\cref{sec:conclu}.

\section{Background}
\label{sec:background}
In this section, we first give preliminaries on unstable features of the Rust compiler and then discuss the cargo package manager and dependency management.

\subsection{Unstable Feature of Rust compiler}
\label{Unstable Feature of Rust compiler}

Rust compiler has three release channels: nightly, beta, and stable~\cite{rustc:channel}. Developers can use any channel releases to build their projects. Especially, the nightly channel provides unstable features to extend the functionalities of the compiler. We define these features as \textbf{Rust Unstable Features} (short for \textbf{RUF}) \cite{RustUnstableBook}.
By adding code \texttt{\small{\#![feature(feature\_name)]}}, developers can enable the RUF \code{feature_name}. 
Rust defines \texttt{\small{\#![feature(feature\_name)]}} as \textbf{RUF configuration}~\cite{rustc:con_compi}.
\cref{code:unstable_feature_example} gives an example of RUF usage. Without the RUF configuration of \texttt{\small \#![feature(box\_syntax)] }, \texttt{\small let x = box 1} will cause a compilation failure as \code{box} cannot be resolved. 
Moreover, developers can also specify \textbf{configuration predicates} (like \code{compiler_flag} or \code{target_os = "linux"}) in the RUF configuration to allow conditional compilation~\cite{rustc:con_compi}, such as specifying operating systems, as shown in~\cref{code:unstable_feature_cfg_example}. 
RUF will be enabled if the configuration predicate is true.

\begin{figure}
    \centering
    \begin{subfigure}[b]{0.4\linewidth}
        \begin{lstlisting}[language=rust, escapeinside={(*@}{@*)}]
#![feature(box_syntax)]
fn main() {
    let x = box 1;
}
        \end{lstlisting}
        \caption{RUF Usage.}
        \label{code:unstable_feature_example}
    \end{subfigure}
    \hfill
    \begin{subfigure}[b]{0.59\linewidth}
        \begin{lstlisting}[language=rust, escapeinside={(*@}{@*)}]
#![cfg_attr(compiler_flag, feature(ruf))]
#![cfg_attr(target_os = "linux",
    feature(llvm_asm))]
    (*@\quad@*)
        \end{lstlisting}
        \caption{RUF Configuration.}
        \label{code:unstable_feature_cfg_example}
    \end{subfigure}
    \caption{RUF example.}
\end{figure}

\noindent \textbf{RUF Status.} RUF have two major types, language features and library features. The language feature is implemented in the Rust compiler to provide compiler support, such as syntax extension, user-defined compiler plugin, etc. 
Library feature is implemented in the Rust standard library to provide extra functionalities that are under development. 
Language feature has four types of statuses: 
\begin{itemize}[noitemsep, leftmargin=*]
    \item \textit{Accepted.} The RUF is stable and integrated into the stable compiler.
    \item \textit{Active.} The RUF is under development and can only be used in a nightly compiler.
    \item \textit{Incomplete.} The RUF is incomplete and not recommended to use. 
    \item \textit{Removed.} Not supported by the compiler anymore.
\end{itemize}

Library feature only has two statuses: \textit{stable} and \textit{unstable}, corresponding to \textit{accepted} and \textit{active}, respectively.

\vspace{+2pt}
\noindent \textbf{RUF Impacts via Dependencies.}
RUF not only impacts Rust packages directly but also affects other packages through dependencies. 
Packages in the Rust ecosystem often reuse other packages, thus propagating RUF impacts. 
%
For example, \code{redox_syscall-0.1.57} uses removed RUF \code{llvm_asm} to implement syscall wrappers~\cite{pack:redox_syscall}, leading to compilation failures.
Even worse, any packages depending on \code{redox_syscall-0.1.57} will get the same compilation failure. This reveals that RUF threats can be amplified through package dependencies in the ecosystem.

\vspace{+2pt}
\noindent \textbf{RUF Threats.} 
\label{ruf_threats}
RUF introduces at least three potential threats to the Rust ecosystem.
\textbf{1) Compilation failure.} Once the RUF is removed, all packages enabling it can no longer compile. As RUF provides functionality extensions at the compiler level, removing or replacing RUF to avoid compilation failure is not easy. 
\textbf{2) Unstable functionality.} RUF are unstable and are still under development. Therefore, the functionalities, syntax, and implementation of RUF can be changed, which may cause unstable behaviors, compilation failures, or even runtime errors~\cite{cve:error_type_id}. 
Even worse, RUF can impact the Rust compiler architecture if implemented poorly, even though it's not enabled~\cite{unnamed_fields_bug}.
\textbf{3) Unstable compiler selection imposed by RUF.} RUF force developers to use nightly compilers. As a result, all projects that depend on RUF must be compiled using nightly compilers, though developers are recommended to use stable ones.

\subsection{Cargo Package Manager and Dependency Management}
\label{Cargo Package Manager and Dependency Management}

\textit{Cargo} is the official package manager of Rust, which manages Rust package dependencies and collaborates with the Rust compiler to build Rust projects. 
It can also be used by developers to upload or download packages to/from the official package registry \code{crates.io}~\cite{cratesio}, which hosts packages for the Rust ecosystem.
\textit{Cargo} uses semantic versioning specifications to manage versions of Rust packages. The basic version format is \code{<MAJOR.MINOR.PATCH>}. 
\code{MAJOR} is increased when there are incompatible API changes.
When there are backward-compatible changes of functionalities added, \code{MINOR} is increased.
\code{PATCH} is increased only when bug fixes are introduced.

\begin{figure}
    \centering
    \begin{subfigure}[b]{0.5\linewidth}
        \begin{lstlisting}[language=rust, escapeinside={(*@}{@*)}]
[dependencies]
rand = { version = "0.1.2",
    optional = true }
[features]
pf = ["dep:rand"]
    (*@\quad@*)
        \end{lstlisting}
        \caption{Dependency requirements}
        \label{code:dep_req}
    \end{subfigure}
    \hfill
    \begin{subfigure}[b]{0.49\linewidth}
        \begin{lstlisting}[language=rust, escapeinside={(*@}{@*)}]
#![cfg_attr(feature = "pf",
    feature(box_syntax))]
#[cfg(feature = "pf")]
pub fn get_box() -> usize {
    return box rand()
}
        \end{lstlisting}
        \caption{PF usage}
        \label{code:pf_usage}
    \end{subfigure}
    \caption{PF example.}
    \label{fig:pf}
\end{figure}

\label{Cargo Dep Type}
\vspace{+2pt}
\noindent \textbf{Dependency Types.} \textit{Cargo} has five types of dependencies \cite{cargo:dep}:
1) Normal dependency. It is used in runtime, which is the most common dependency type.
2) Build dependency. It is used in building scripts to create environments like data, code, etc.
3) Development dependency. It is used only in tests, examples, and benchmarks.
4) Target (Platform-specific) dependency. Dependency marked \textit{target} is enabled only in a specific platform like Windows or other platforms, according to its definition.
5) Optional dependency. Dependency marked \textit{optional} is not enabled by default. This type is enabled only when related \textit{package feature} is enabled.

\label{package_feature}
\vspace{+2pt}
\noindent \textbf{Package Feature (PF).} Rust packages can also define \textit{features} to allow conditional compilation, which is related to optional dependencies \cite{cargo:feature}.
To distinguish from RUF, we define these \textit{features} as Package Feature (PF).
Codes marked with PF are compiled only when the PF is enabled.
Unlike RUF, PF is implemented by developers rather than the Rust compiler.
PF is usually tied with specific functionality of packages to satisfy different requirements of developers using the package.
\textit{Cargo} allows developers to bind PF with optional dependencies. When PF is used, related optional dependencies are enabled to support the implementation of PF codes.

Take \cref{fig:pf} for example, in which we define \code{pf} that depends on \code{rand} (last line of \cref{code:dep_req}), as the implementation of \code{pf} needs function \code{rand()} from package \code{rand}.
In \cref{code:pf_usage}, when \code{pf} is enabled, \code{get_box()} is compiled, and optional dependency \code{rand} is introduced into the package.
As the implementation of \code{get_box()} also needs the RUF \code{box_syntax}, we use configuration predicates to declare that the RUF \code{box_syntax} is enabled when \code{pf} is enabled.
It is worth mentioning that this example also shows that if other packages use this with \code{pf} disabled, the RUF is disabled, too.

\section{RUF Status and Usage Extraction}
\label{RUF Extraction}

To conduct RUF usage and impacts study for the Rust ecosystem, we first need to extract RUF definitions, status, and usages.
The RUF extraction process consists of two steps, as shown in \cref{fig:impl_arch}.
First, we extract and track all RUF that the Rust compiler supports to reveal RUF status changes over time.
Second, we develop two new techniques to extract and understand RUF usages in Rust packages.

\subsection{RUF Status Extraction}
\label{RUF Status Tracking}

As discussed in \cref{Unstable Feature of Rust compiler}, RUF status changes between different compiler versions. An \textit{active} RUF in a specific compiler version may get stabilized or removed in the following compiler versions, depending on its development process.
To investigate RUF evolution over time, we need to resolve several technical challenges to track the status of RUF among all compiler versions over time.
First, there is no official documentation on RUF. All RUFs are defined in the Rust compiler source code.
Second, the syntax of RUF is not unified. Library features and language features of RUF have different definition syntaxes. What's worse, the definitions of RUF are scattered in different locations, which makes it hard to track them.
Third, the Rust compiler frequently changes its architecture, causing RUF definition syntax to change between compiler releases.
The Rust compiler provides \textit{tidy} \cite{rust:tidy}, which can be used to track RUF status.
However, it only covers partial RUF definition and does not support old Rust compiler versions.

To conquer the above challenges, we design and implement our RUF status tracker to detect all RUF. 
To extract the language feature, we observe that its definition syntax can be described as \code{(RUFStatus, RUFName, OtherAttributes)}. However, the order of each attribute in the definition and supported attributes may change in different versions of the compiler.
Therefore, we use two regular expressions to match all types of syntax change during compiler release update, including  \code{("([a-zA-Z0-9]+?)", .+,(Active|} \code{Accepted|Removed))} and \code{((active|accepted|removed),([a-zA-Z0-9} \code{]+?),.+)}.
To extract library features, 
we use \textit{tidy} to recognize each attribute in the definition and then merge them to form complete RUF status information. We extend \textit{tidy} to detect complete library feature definitions in all release versions of the compiler.
To ensure both accuracy and coverage, we include all Rust source code files in the Rust compiler but exclude test-related files.

Besides RUF definition parsing, we further detect abnormal RUF status transitions to explore the gap between ideal and real-world RUF development. We detect three types of abnormal transitions:
1) \textit{Accepted} RUF change to any other status. This is abnormal as stabilized RUF should not return to unstable status.
2) \textit{Removed} RUF change to any other status.
3) RUF supported by the old Rust compiler are not recognized by newer compiler versions. 
By detecting abnormal RUF status transitions among compiler release updates, we conduct an in-depth study of RUF lifetime, discussed in~\cref{RQ1}.

\subsection{RUF Usage Extraction}

\subsubsection{RUF Configuration Extraction}
\label{RUF Usage Extraction}

Aside from extracting all RUF that the compiler supports, we further investigate RUF usage in the ecosystem. To achieve this, we have to extract \textbf{RUF configurations} (details in \cref{Unstable Feature of Rust compiler}) used by Rust developers in the package.
We need to extract the RUF configuration and its enable condition in each Rust package in the ecosystem, which requires both accuracy and efficiency.
However, RUF configurations can be defined in complex syntax, so regular expressions cannot guarantee both accuracy and coverage. 
Although the compiler can accurately recognize RUF configurations defined in packages, the full compilation of all packages takes unbearable time. What's worse, the compilation process only resolves configuration predicates with user-given options and local runtime environments. 
This makes it almost impossible to cover all possible compilation conditions and will lose the coverage of RUF usage.  

\begin{figure}[t]
	\centering
	\includegraphics[width=0.9\linewidth]{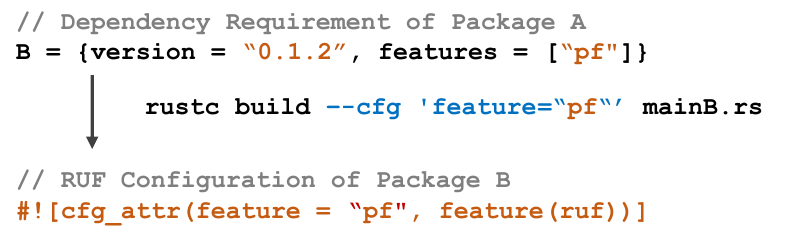}
    \caption{RUF impact example.}
    \label{fig:ruf_impact}
\end{figure}

To resolve these challenges, we propose a \textit{compilation data-flow interception-based RUF configuration extractor}. 
For compatibility, we integrate our extractor into Rust compiler compilation options. When developers specify the option, the compiler will start extracting RUF configurations. 
To ensure efficiency, instead of compiling the whole package, we exclude package configurations and source codes.
Moreover, we only analyze configuration predicates in library files, where RUF are used, according to Rust's official documents. After collecting the necessary compilation data, we intercept the data flow and redirect it to our extractor.
Reusing the query system of the Rust compiler, we acquire all configurations and filter out RUF-related ones.
We avoid additional configuration processing to get original data, and then parse configuration predicates.
When the process is done, we terminate the compilation process immediately and will not generate any compilation file to reduce I/O operations.
The implementation result shows that we can successfully extract RUF usage from over \data{99.6\%} of the Rust package versions in the ecosystem. The rest are caused by Rust syntax incompatibility with old packages.
%


\subsubsection{Semantic Identification of RUF Configuration}
\label{Semantic Identification of RUF Configuration}

Although we have collected all RUF configurations in the Rust ecosystem, the RUF configuration predicates are formatted as strings in the compiler without semantics. 
To decide whether the RUF is enabled, we must identify the semantic relations between configuration predicates and dependency requirements.
Using code in~\cref{fig:ruf_impact} as an example, package B defines RUF configuration with the predicate \code{pf}, which means that \code{ruf} is enabled only when \code{pf} of package B is enabled. In the compilation process of package A, the Rust compiler receives compiler flag \code{--cfg `feature="pf"'} and determines the RUF configuration predicate is satisfied. In this case, \code{ruf} is enabled.

Whether the RUF is enabled can only be determined in the compilation process. As a result, an intuitive solution is to compile package A to get its information.
However, the compilation of all Rust packages takes unbearable time. 
What's worse, developers can use keywords \code{}{All/Any/Not} to define nested predicates like \code{ALL(ANY(linux, target_env = "sgx"), feature = "pf")}. In this case, we need to explicitly specify compiler flags (\code{linux} and \code{sgx}) in the compilation process. Otherwise, our analysis will assume that the RUF is disabled, leading to an inaccurate RUF impact analysis.

To accurately determine RUF impacts, we propose the \textit{semantic identification of RUF configuration}. 
The basic idea is to split RUF configuration predicates into minimal compiler flags, and identify the semantic relationship between the flags and dependency requirements. In this way, we can accurately determine RUF impacts in generated EDG (discussed in~\cref{RUF Impact Resolution}) without compilation. 
%
We use an example of RUF configuration predicates \texttt{\small{ALL(ANY(A,B),C)}} to explain our design. First, we recursively resolve nested configuration predicates. After that, \texttt{\small{ALL(ANY(A,B),C)}} is resolved to be \texttt{\small{[AC, BC]}}. The RUF configuration is then formatted into {\small{$v \xrightarrow[]{AC} RUF$}} and {\small{$v \xrightarrow[]{BC} RUF$}}. 
The {\small{$v \xrightarrow[]{AC} RUF$}} means the RUF is enabled in version $v$ when predicates $AC$ ($A\wedge C$) is satisfied.
After that, we define \textbf{corpus function} $\delta(dep, cfg)$, which satisfies $\delta(dep, AC) = \delta(dep, A) \wedge \delta(dep, C)$.
$\delta(dep, cfg)$ is true when dependency $dep$ satisfies the RUF configuration predicates of $cfg$.
Using the corpus function, we can determine whether the dependency $dep$ will enable RUF.

We build corpus function according to \textit{Cargo} dependency definition syntax and its relation with the compiler flags transferred to the Rust compiler based on official documentation \cite{cargo:dep, cargo:feature, rustc:con_compi, rustc:rustc-dev-guide}.
There are 15\% of predicates that are not officially documented.
7\% are obvious community conventions (e.g., "docs"). 
To ensure correctness, we only include obvious conventions. We also randomly select packages with these predicates and compile them to make sure that the conventions are all followed.
%
Other 8\% of predicates are hard to find such obvious conventions, and we assume they will not impact other packages by default. In this way, we may underestimate the RUF impact.

\begin{figure}
    \centering
    \includegraphics[width=\linewidth]{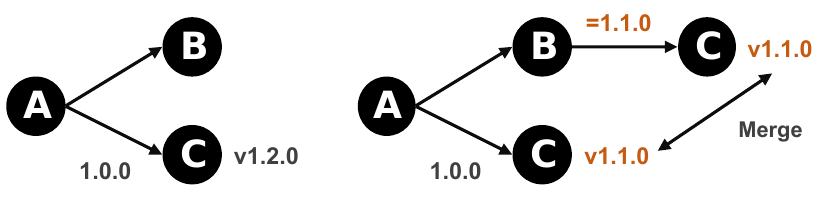}
    \caption{Dependency influence example. While A doesn't change its dependency requirements of C, the other dependency from B to C will influence it and result in choosing a different version of C.}
    \label{fig:dep}
\end{figure}

\section{Quantitative Analysis of RUF Impacts}
\label{RUF Impact Resolution}

To quantify RUF impacts over the whole Rust ecosystem, we first define \textbf{E}cosystem \textbf{D}ependency \textbf{G}raph (EDG) and factors that affect impact propagation (\cref{RUF Impact Definition}).
We further propose a new technique to accurately and efficiently resolve dependencies in the entire ecosystem (\cref{Dependency Resolution}).
After EDG is generated, we can determine RUF impacts in the ecosystem.
The evaluation results show that the proposed dependency resolution technique can achieve 99\% accuracy(\cref{Accuracy Evaluation}).

\subsection{EDG and RUF Impact Definition}
\label{RUF Impact Definition}

Packages can specify dependency requirements to declare what packages they need. After the dependency resolution process of a package, the dependent package versions and attributes are determined. We define these resolved dependencies as the dependency tree (DT) of the package.
We define \textbf{EDG} as $G=(N,E)$, where each node $v$ in $N$ represents each Rust package version, and each edge $dep$ in $E$ represents transitive dependency between nodes, including both direct and indirect dependencies.
Direct dependency is defined as format $v_a \xrightarrow[]{dir} v_b$, where $v_a$ directly depends on $v_b$. We also define indirect dependency format $v_a \xrightarrow[]{indir} v_b$, where $v_a$ indirectly depends on $v_b$. We define \textit{dep} relation $v_a \xrightarrow[]{dep} v_b$ if version $v_b$ is in $DT_a$, where $v_a$ transitively depends on $v_b$.

\begin{figure}
    \centering
    \includegraphics[width=\linewidth]{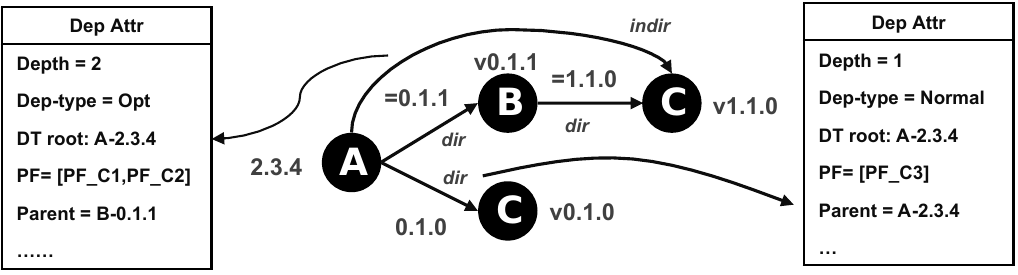}
    \caption{EDG structure example.}
    \label{fig:edg}
\end{figure}

A simple but inaccurate method to generate EDG is to resolve all direct dependencies and connect them. In this case, when package $p_a$ depends on $p_b$ and $p_b$ depends on $p_c$, then $p_a$ depends on $p_c$. However, this is not accurate for Rust.
For example, if a specific package version $v_a$ directly depends on another version $v_b$ in the DT of $v_a$ ($DT_a$) and $v_b$ directly depends on $v_c$ in $DT_b$, we can't guarantee $v_a$ transitively depends on $v_c$ in $DT_a$.
This is because dependencies can influence each other, as shown in \cref{fig:dep}. Compatible dependencies may be merged, which makes the same dependency requirements choose different dependent versions.
In \cref{fig:dep}, although A does not change its dependency requirements of C, after the dependency resolution of B, the final choice of C changes as influenced by dependency from B to C.
As a result, it is compulsory to resolve all dependencies before we get DT. Every dependency change needs a complete resolution process to generate a new DT.
Aside from versions, package features (PF), dependency type, and other dependency attributes influence the resolution process.
In this way, we need to store extra attributes in the EDG $dep$ to determine dependencies accurately, as shown in \cref{fig:edg}.

To further determine RUF impacts through EDG, we first define RUF data structures.
In \cref{RUF Usage Extraction}, we get RUF configurations in each Rust package. We define \textbf{RUF usage set $T$ }that contains all of these configurations ($cfg$ for short). Each element inside $T$ is formatted as $v \xrightarrow[]{cfg} RUF$, which means package version $v$ enables $RUF$ when the configuration predicates of $cfg$ are true. And the \textbf{corpus function $\delta(dep, cfg)$} defined in \cref{Semantic Identification of RUF Configuration} determines whether dependency $dep$ satisfies the RUF configuration predicates of $cfg$.

The RUF impacts can be divided into direct and transitive ones:
1) Direct Impact (Equation \ref{equa:dir_rufimpact}): Packages that directly use RUF are impacted by RUF.
2) Transitive Impact (Equation \ref{equa:indir_rufimpact}): Packages that transitively use RUF by DTs are impacted by RUF. To get the indirect RUF impact of given $RUF$, we first find every $v_b$ using it. Then, we select all package versions $v_a$ that depend on $v_b$ directly or transitively. Last, we judge whether its dependency satisfies the RUF configuration predicates using our corpus function.
The impact definition of certain RUF or a category of RUF is similar, as we only consider corresponding RUF in $T$ rather than all of them. 

\begin{equation}
     DirImpact(RUF) = \{v \in N |  (v \xrightarrow[]{cfg} RUF) \in T \}
\label{equa:dir_rufimpact}
\end{equation}

\begin{equation}
\begin{split}
    TranImpact(RUF) = \{v_a &\in N  |  \exists v_b ((v_a \xrightarrow[]{dep} v_b) \in E \ \wedge
    \\ & ( (v_b \xrightarrow[]{cfg} RUF) \in T \wedge \delta (dep, cfg)) \}
\end{split}
\label{equa:indir_rufimpact}
\end{equation}

\subsection{Ecosystem Dependency Resolution}
\label{Dependency Resolution}

To quantify RUF impacts, we need to generate EDG by resolving dependencies of all Rust packages. Although \textit{Cargo} provides an official dependency resolution tool, it lacks flexibility and takes unbearable time to resolve the entire ecosystem. 
To achieve both accuracy and efficiency, we face several specific challenges.
Sampling for ecosystem analysis\cite{py_dep, java_dep} lacks coverage.
Assuming that all dependencies are transitive and ignore package-manager-specific rules \cite{small_world, npm_vul_impact} gains coverage but loses accuracy.
Existing work simulating the resolution rules increases accuracy, but they fail to cover all dependency types and only use an approximate resolution strategy, making the resolution less accurate~\cite{npm_vul}.
Moreover, as RUF impacts other packages conditionally, our resolver should be able to resolve and store RUF-related configurations (e.g., PFs) aside from dependency. Otherwise, RUF impacts cannot be precisely determined.
%
%

\begin{figure}[t]
\begin{algorithm}[H]
    \caption{Rust Ecosystem Dependency Graph Generation}
    \label{alg:cap}
    \begin{algorithmic}[1]
        \Require $N$ - Unresolved Package Versions
        \Ensure $G$ is Ecosystem Dependency Graph (EDG)
        \State $E \gets \emptyset$
        \State $resolver \gets new \ VirtualCargoEnv()$
        \ForEach{$ v \in N$}
            \State $pf_v \gets v.all\_pf()$
            \State $cfg \gets new \ VirtualPackConfig(v, pf_v)$
            \State $cfg \gets cfg.dep(normal \cup build \cup opt \cup target)$
            \State $res \gets  resolver.resolve(cfg) $
            \ForEach{ $v_i \in res.ver$ }
                \ForEach{ $(v_i \xrightarrow[]{dir} v_j) \in res.dir(v_i)$ }
                    \State $attr \gets format(v, dir, v_i, v_j) $
                    \State $dep_{j} \gets new\ Dep(v \xrightarrow[]{attr} v_j)  $
                    \State $E \gets E \cup dep_{j}$
                \EndFor
            \EndFor
        \EndFor
        \State $G \gets (N, E)$
        \State
        \Return $G$
    \end{algorithmic}
\end{algorithm}
\end{figure}
 
To conquer these challenges, we design our own EDG generator to resolve dependencies.
To achieve accuracy, we use the \textit{Cargo} core resolver to resolve dependencies. However, the core resolver needs \textit{Cargo} and package environment to analyze the dependency requirements defined in the Rust packages. The default environment provided by \textit{Cargo} is not extensible as the process is blocked by I/O operations and only allows single thread execution.
To conquer this problem, we build a virtual environment for the core resolver and virtualize package configuration before dependency resolution.
The virtual package environment only contains the minimal configuration of the target package to be resolved.
The resolution process uses its own \textit{Cargo} environment rather than the one shared in the host machine. Thus we can extend the resolution process in any threads and avoid locks. 

EDG generation process can be described by \cref{alg:cap}. We first create Virtual Environment Configuration (line \data{2} and \data{5}). After that, we adjust our package configuration with all package features (PF) and all dependency types except development dependency enabled. This is because development dependency is only used for tests, examples, and benchmarks, thus not impacting package runtime, mentioned in \cref{Cargo Dep Type}. After that, we resolve the DT of $v$ to get results $res$. It is then formatted into EDG edges (lines \data{7-14}).
The key EDG format process $format(v, dir, v_i, v_j)$ (line \data{10}) is designed for both efficiency and flexibility.
We can only keep necessary dependency information like PFs. This accelerates the resolution process and reduces the EDG size. Moreover, the format process is user-defined and gives more possibility to include other dependency attributes for further investigation other than RUF impact.

In our implementation, we build a channel-based resolution pipeline.
The sender packs unresolved package version information and sends it to the receiver. The receiver is responsible for the resolution process described in \cref{alg:cap}.
The virtual environment configuration is achieved by setting environment variables of \textit{Cargo}, which creates a virtual workspace for it. We virtualize the package environment by constructing a fake configuration file containing target dependency, type, PF, and other attributes according to version metadata. 
In the format process, we only store transitive dependencies linked to the root version without any attributes for efficiency.
After constructing EDG, we resolve dependencies with attributes a second time using the same pipeline. This time, we only resolve necessary versions and dependencies according to RUF configurations to get RUF impacts.
The EDG generation process takes only about \data{2 days} to resolve all \data{140M} transitive dependencies in an AlderLake machine, and EDG occupies only \data{2 GB} of storage. Based on EDG, our ecosystem-level analysis only takes seconds or minutes to complete, with all transitive dependencies counted.


\subsection{Accuracy Evaluation}
\label{Accuracy Evaluation}

As different package manager uses different dependency resolution strategy, there is no standard resolution accuracy benchmark. For evaluation, we select \textit{Cargo Tree} tool from the official package manager \textit{Cargo-1.63.0} for comparison. We resolve four types of dependency: build, common, optional, and target. Only development dependencies are omitted as they will not affect the runtime of programs, which is the same as our resolution rules.
We first download source code from the official database \textit{Crates.io}, and then use \textit{Cargo Tree} to resolve dependencies in the real environment. After that, we will compare dependency items from \textit{Cargo Tree} and our ecosystem dependency graph.
To evaluate accuracy in the different data sets, we select 2000 packages from the whole ecosystem as our standard dependency benchmark data set and choose three strategies to select these packages:
1) Random: Randomly selected versions.
2) Popular: Latest versions of packages that have the most downloads.
3) Mostdep: Latest versions of packages that have the most direct dependencies, which is the most complex situation a resolver will meet.
We select the latest versions of the given package because it is chosen to be the dependency package version by default and typically has the most complex dependencies.

We define four types of comparison results given package \textit{i} in the accuracy evaluation: 
1) $Right$ ($R_i$): Dependencies that occur in both dependency data sets with the same versions. 
2) $Wrong$ ($W_i$): Dependencies that occur in both dependency data sets with different versions. 
3) $Over$ ($O_i$): Dependencies that only occur in our resolution data set.
4) $Miss$ ($M_i$): Dependencies that only occur in standard data sets.
We treat each dependent version as a dependency and the sum of dependent versions as a dependency tree in the evaluation process. This is because we only care about whether a specific package version impacts the root package in the dependency tree rather than how it impacts the dependency tree.

\sloppy We use four indexes to represent accuracy shown in \Crefrange{equa:TreeAcc}{equa:F1score}.
 $Tree Accuracy$ stands for the resolution accuracy of the entire dependency tree.
 $Recall$ and $Precision$ represent $Right$ percentage in standard dependency and resolved dependencies data set, respectively.
$F_{1}-score$~\cite{wiki:F-score} is the harmonic mean of recall and precision, which can represent the accuracy of resolution. 

\vspace{+2pt}
\begin{equation}\label{equa:TreeAcc}
         Tree Accuracy = \frac{\sum_{i=1}^n [W_i + O_i + M_i = 0]}{n} 
\end{equation}
\begin{equation}
         Recall = \frac{\sum_{i=1}^n R_i}{\sum_{i=1}^n (R_i+M_i)}
\end{equation}
\begin{equation}
         Precision = \frac{\sum_{i=1}^n R_i}{\sum_{i=1}^n (R_i+W_i+O_i)}
\end{equation}
\begin{equation}\label{equa:F1score}
         F_{1} - Score = \frac{2*Recall*Precision}{Recall+Precision}
\end{equation}
\vspace{+2pt}

\begin{table}[]
    \caption{\data{Resolution accuracy}.}
    \label{tab:accuracy}
    \begin{tabular}{ccccc}
    \hline
    Dataset Type & Tree Accuracy & Precision & Recall  & F1Score \\ \hline
    Random       & 99.39\%       & 99.76\%   & 98.88\% & 99.32\% \\
    Popular      & 99.50\%       & 99.99\%   & 99.16\% & 99.58\% \\
    Mostdep      & 97.04\%       & 99.96\%   & 97.47\% & 98.70\% \\ \hline
    \end{tabular}
\end{table}

Results in \cref{tab:accuracy} show that our dependency resolution tool can achieve \data{97.04-99.50\%} accuracy in dependency tree resolution. The EDG structure has \data{99.76-99.99\%} precision, which means the dependencies in EDG are mostly accurate. Moreover, EDG has \data{97.47-99.16\%} recall, which shows it only loses a tiny number of dependencies from the Rust ecosystem.
In the evaluation process, we observed that the dependency configuration behaves slightly differently when it is uploaded to the ecosystem rather than built locally. The configuration file in the source code may force developers to use a specific version of the package manager, resolver, or compiler during the local development of the built package.
Furthermore, it will probably use local packages instead of packages from \textit{Crates.io}. These operations are forbidden when they are uploaded to \textit{Crates.io} and used by other packages. This configuration setting is mainly used for local environments but not for other developers who want to use the functionalities of this package. As a result, our evaluation process removes local configurations to keep consistent with the Rust ecosystem behavior.

\section{Ecosystem-scale Study}
\label{Ecosystem-scale Study}

Our source data comes from the official package database \textit{crates.io} on \data{August 11, 2022}, which contains \data{592,183} package versions.
In total, we resolve \data{139,525,225} transitive dependencies and extract \data{182,026} RUF configurations. 
To drive our study on RUF, we raise three research questions (RQs).
\begin{itemize}[noitemsep, leftmargin=*]
  \item \textbf{RQ1}: (RUF Lifetime) How does Rust Unstable Feature status evolve with time?
  \item \textbf{RQ2}: (RUF Usage) How do packages in the Rust ecosystem use Rust Unstable Feature?
  \item \textbf{RQ3}: (RUF Impacts) How does Rust Unstable Feature impact packages in the Rust ecosystem?
\end{itemize}

\subsection{RQ1: RUF Lifetime}
\label{RQ1}


By tracking RUF status in every minor version of the Rust compiler from \data{v1.0.0 (2015-05-15)} to \data{v1.63.0 (2022-08-11)}, we obtain \data{1,875} RUF supported by Rust compiler, including both language features and library features. In the latest version of the compiler, RUF status is shown as follows: \data{\textit{Accepted}(1,002), \textit{Active}(562), \textit{Incomplete}(11), \textit{Removed}(59), \textit{Unknown}(241)}. 
Typically, new RUF will first appear as \textit{active} or \textit{incomplete} status and will be stabilized to \textit{accepted} status after RUF development. During the RUF status evolution, the RUF may be judged useless and get \textit{removed}.
Half of RUF (\data{47\%, 873/1,875}) are not stabilized in the latest version of the Rust compiler, and \data{16\%} of RUF (\data{300/1,875}) development eventually stops and becomes \textit{removed} or \textit{unknown}. 


As discussed in \cref{RUF Status Tracking}, we further detect abnormal RUF status transitions to reveal unexpected RUF development behavior. We observe \data{277/1,875 (15\%)} abnormal RUF status transitions.
Among these abnormal RUF, \data{244} RUF supported by the old compiler are not recognized by the new compiler, thus becoming \textit{unknown}. Moreover, \textit{accepted} (i.e., stable) RUF can become \textit{active} or even \textit{unknown} during development. Some RUF appear and disappear repeatedly and eventually fail to be stabilized.
After inspecting abnormal RUF status evolution, we find that the implementation of some RUF (e.g., \code{unnamed_fields} ~\cite{unnamed_fields_bug}) interferes with the architecture of the Rust compiler and makes it produce unexpected behavior.
The RUF \code{unnamed_fields} breaks the reliability of the compiler. This makes the compiler unable to recognize correct programming syntax even though the RUF is not enabled. 
There are other reasons for RUF removal other than bugs. The functionality of the RUF may overlap with other stable methods or existing RUF, so the RUF is considered useless and gets removed. But the removal causes incompatibility of RUF usage, thus introducing compilation failure.
We suggest introducing another status that marks the RUF as \textit{inactive}. If the RUF is not proven to be vulnerable or unstable, it can still work for developers, but with a warning instead.
\begin{flushleft}
    \fbox{\begin{minipage}[l]{.462\textwidth}
    \textbf{Finding-1:} 
    We observe \data{277/1,875 (15\%)} abnormal RUF status transitions, which are mainly caused by abandoned development or bugs. RUF may get removed and cause package compilation failures, even though they are not vulnerable to packages. These types of abnormal RUF lifetime break the usability of Rust packages.
    
    
    \end{minipage}}
\end{flushleft}

Although half of RUF are stabilized in the latest version of the Rust compiler, there is still potential instability in the \textit{accepted} RUF.
We further focus on \textit{accepted} RUF that returns to unstable status. It indicates that there may be instability found in the stable RUF.
While \textit{accepted} RUF can be integrated into a \textit{stable} Rust compiler, this does not mean it is completely safe for packages that use it.
While RUF \code{proc_macro} \cite{ruf:proc_macro} is widely used in the Rust ecosystem, the stabilization process is tough as it turned to \textit{accepted} in \code{v1.15.0} of the Rust compiler and then came back to \textit{active}.
RUF \code{error_type_id} was accepted in \code{v1.34.0} of Rust compiler, but then its implementation was found not memory safe~\cite{cve:error_type_id}, and it returned to \textit{active}.
This means that stabilized RUF can also introduce security threats to packages. Moreover, the stabilization process of RUF needs to be carefully discussed, especially when deciding whether the RUF should change its status or not.


\begin{flushleft}
    \fbox{\begin{minipage}[l]{.462\textwidth}
    \textbf{Finding-2:} We find that half of RUF (\data{47\%}) are not stabilized in the latest version of the Rust compiler. Even worse, stabilized RUF can also introduce vulnerabilities to the package \cite{cve:error_type_id}, which indicates that stabilized RUF cannot be regarded as totally safe for developers. 
    \end{minipage}}
\end{flushleft}



\subsection{RQ2: RUF Usage}
From all package versions of the Rust ecosystem, we extract \data{1,000} RUF and \data{182,026} RUF configurations used by  \data{72,132 (12\%)} versions in total. It is worth mentioning that although the Rust compiler supports \data{1,875} RUF, not all of them are used by Rust packages.
Our implementation of Rust Unstable Feature Extraction is based on \data{Rustc-v1.63.0-Nightly}.
Though our study only includes packages that could be pre-compiled in the latest version of the Rust compiler, we still use all packages in the ecosystem as complete works.

\begin{table}[]
    \caption{\data{Summary of RUF usage.}}
    \label{tab:ruftype}
    \resizebox{0.9\columnwidth}{!}{%
        \begin{tabular}{lcrr}
        \hline  
        Type       & \multicolumn{1}{l}{RUF Count} & \multicolumn{1}{l}{Package Versions} & \multicolumn{1}{l}{RUF Usage Items} \\ \hline
        Accepted   & 382 (38\%)                    & 24,681 (34\%)                         & 38,858  (21\%)                       \\
        Active     & 381 (38\%)                    & 55,785 (77\%)                         & 101,494 (56\%)                       \\
        Incomplete & 7   (7\%)                     & 5,829 \ \ (8\%)                          & 5,926 \ \ (3\%)                        \\
        Removed    & 41  (4\%)                     & 14,812 (21\%)                         & 21,096  (12\%)                       \\
        Unknown    & 189 (19\%)                    & 10,534 (15\%)                         & 14,652 \ \ (8\%)                        \\
        Total      & 1000(100\%)                   & 72,132(100\%)                        & 182,026(100\%)                      \\ \hline
        \end{tabular}
    }
\end{table}

~\cref{tab:ruftype} shows RUF usage status in the Rust ecosystem.
Only \data{38}\% types of RUF become stable while only accounting for \data{21}\% of usages. There are still \data{38}\% types of RUF that are \textit{active} and need further development. What's worse, \data{23}\% types of RUF are \textit{unknown} or \textit{removed}. Packages that enable such RUF can't be compiled.
The present situation of RUF usage in the Rust ecosystem is even worse. \data{72,132 (12\%)} package versions are using RUF, and \data{65,172/72,132 (90\%)} package versions among them are still using unstabilized RUF, which means using RUF that is not \textit{accepted}. This indicates that unstabilized RUF usage still dominates among all RUF. In addition, \data{21,338/72,132 (30\%)} of these package versions are using \textit{removed} or \textit{unknown} RUF, which makes them directly suffer from compilation failure.

We also conducted limited research on large projects of RUF usage, including the Android Open Source Project (AOSP), the Linux operating system, and the Firefox browser.
These projects are all widely used and have strong requirements for reliability.
We discovered that they all used RUF in their main repository \cite{AOSPRUFUsage, FirefoxRUFUsage, LinuxRUFUsage, LinuxRUFUsage2}.
Moreover, AOSP and Firefox also cloned the source code from third-party packages to their main repository \cite{FirefoxRUFUsage, AOSPRUFUsage3Party}, which introduced extra RUF usage.
What's worse, there is still \textit{removed} RUF usage integrated into the main repository \cite{android_ruf}, which could break the reliability and usability the project. For stabilization, the RUF usage should be carefully reviewed and discussed to make sure that it won't cause reliability issues or vulnerabilities.
\begin{flushleft}
    \fbox{\begin{minipage}[l]{.462\textwidth}
    \textbf{Finding-3:} Although RUF are declared to be experimental extensions and unstable for Rust developers, \data{72,132 (12\%)} package versions in the Rust ecosystem are using RUF, and \data{90\%} of package versions among them are still using unstabilized RUF.
    
    \end{minipage}}
\end{flushleft}

\subsection{RQ3: RUF Impacts}
 We resolve dependencies from all package versions in \textit{crates.io} with \data{4,508,479} direct dependencies and successfully collect \data{139,525,225} transitive dependencies from \data{479,201(81\%)} versions. Our implementation of the EDG generator is based on \data{Cargo-1.63 2022-08-11, Resolver V2}.
We do not cover all versions because most unresolved versions have no dependency or have resolution conflicts. The accuracy evaluation in \cref{Accuracy Evaluation} shows that the coverage should be over \data{97-99\%}. While our study only focuses on packages that can be successfully resolved in the newest version of dependency resolver, we still use all packages in the ecosystem as complete works. This means that our data will underestimate the dependency impacts on the ecosystem.

\begin{table}[t]
    \caption{\data{Summary of RUF impacts.} The table shows how package versions are impacted by different types of RUF and through different dependencies.}
    \label{tab:rufimpact}
    \resizebox{\columnwidth}{!}{%
        \begin{tabular}{lrrrr}
        \hline
            RUF Type   & Direct Usage & Uncond Impact& Cond Impact  & Total        \\ \hline
            Accepted   & 24,681        & 17,085        & 21,477        & 38,448 \ (6\%)   \\
            Active     & 55,785        & 77,582        & 207,133       & 237,386(40\%) \\
            Incomplete & 5,829         & 7,696         & 7,991         & 12,097 \ (2\%)   \\
            Removed    & 14,812        & 50,896        & 53,159        & 61,160(10\%)  \\
            Unknown    & 10,534        & 46,157        & 48,742        & 57,916(10\%)  \\
            Total      & 72,154(12\%)  & 111,140(19\%) & 220,665(37\%) & 259,540(44\%) \\ \hline
        \end{tabular}
    }
\end{table}

\label{RUF Impacts Analysis}
~\cref{tab:rufimpact} shows RUF impacts in the Rust ecosystem.
\textit{Direct Usage} represents package versions that directly use RUF, described in Equation \ref{equa:dir_rufimpact}.
\textit{Uncond Impact} represents package versions impacted through dependencies where RUF are enabled by default.
Described in Equation \ref{equa:indir_rufimpact}, \textit{Cond Impact} represents all package versions impacted through dependencies, including \textit{Uncond Impact}.
\textit{Total} represents the total package versions impacted directly or through dependencies.
The results show that, although RUF are used by only \data{72,154 (12\%)} package versions in the Rust ecosystem, it can impact up to \data{259,540 (44\%)} package versions through dependencies, which is almost half of the Rust ecosystem.

As mentioned earlier, RUF are designed to be an unstable extension for preview functionalities. However, it significantly impacts the Rust ecosystem against the original intention of RUF design. 
Among all types of RUF, \textit{unknown} and \textit{removed} RUF can impact a maximum of \data{70,913} package versions, accounting for \data{12\%} of all package versions. This makes them suffer from compilation failures. \textit{Active} RUF affect \data{40\%} of Rust packages in the entire ecosystem, accounting for \data{91\%} of impacted package versions.
At the same time, while there are only \data{8} types of \textit{incomplete} RUF, they affect \data{12,097} package versions. \textit{Incomplete} RUF are extremely unstable, and their API can change at any time, which exposes packages to unexpected runtime behavior and compromises their stability.
\begin{flushleft}
    \fbox{\begin{minipage}[l]{.462\textwidth}
    \textbf{Finding-4:} Through transitive dependencies, RUF can impact \data{259,540 (44\%)} package versions. Removed RUF can cause at most \data{70,913 (12\%)} versions to suffer from compilation failure. This reveals the importance of stabilizing RUF for Rust ecosystem reliability.
    \end{minipage}}
\end{flushleft}

We further analyze why RUF can impact such a large number of packages in the Rust ecosystem. For packages that use RUF, we discover some super-spreaders which many packages depend on.
Taking \textit{unconditional RUF configurations} as an example, we find that \code{redox_syscall-0.1.57} \cite{pack:redox_syscall} uses \textit{unknown} RUF \code{llvm_asm} and \textit{removed} RUF \code{const_fn} and enables them by default. This causes compilation failures for \data{41,750} Rust package versions in the ecosystem, accounting for \data{41,750/46,157 (90\%)} of all package versions unconditionally impacted by \textit{unknown} RUF.
This reveals the great impact of Rust super-spreaders, which is not caused accidentally.
The observed impact is a direct consequence of the centralized Rust ecosystem.
Based on generated dependency graph, we discover a lot of super-spreaders in the Rust ecosystem.
For example, \code{libc-0.2.129} and \code{unicode-ident-1.0.3} have most dependents, \data{353,805 (60\%)} and \data{332,951 (56\%)} package versions respectively. If these packages are affected by \textit{removed} RUF, they will cause massive compilation failures and destabilize the entire ecosystem.
To avoid single-point failure in the ecosystem, super-spreaders are recommended to backport their fix to old versions.
Under the semantic versioning dependency mechanism, the fix can automatically transfer to the whole ecosystem, as the newest version in the compatibility range is usually the first choice of dependencies.

\begin{flushleft}
    \fbox{\begin{minipage}[l]{.462\textwidth}
    \textbf{Finding-5:}
    One of the super-spreaders (\code{redox_syscall-0.1.57}) makes \data{41,750} Rust package versions in the ecosystem fail to compile, accounting for \data{90\%} of unconditionally impacted versions by \textit{unknown} RUF. 
    Once RUF introduce reliability or security problems to super-spreaders, the entire ecosystem could be threatened.
    
    
    \end{minipage}}
\end{flushleft}

\section{RUF Impact Mitigation}
\label{sec:Mitigation}

\begin{table}[]
\caption{An example of RUF recovery that shows a package impacted by RUF A\&B\&C. In 1.57.0, all RUF used are at their best status, so 1.57.0 will be the compatible compiler that can compile the package.}
\label{tab:rufrem_exam}

    \resizebox{0.8\columnwidth}{!}{%
        \begin{tabular}{cccc}
        \hline
        Compiler Version & 1.50.0     & 1.57.0       & 1.63.0  \\ \hline
        RUF A            & Active     & Accepted     & Removed \\
        RUF B            & Accepted   & Accepted     & Accepted\\
        RUF C            & Active     & Accepted     & Unknown \\
        Recovery Status  & \ding{51}  & \ding{51}    & \ding{53}  \\ \hline
        \end{tabular}
}
\end{table}

As our ecosystem-scale study shows that RUF impact a wide scope of Rust packages, it is important to mitigate RUF impacts to avoid potential instability and compilation failure.
In this section, we first discuss our new tool to detect RUF dependency in Rust packages and recover compilation failure caused by RUF impact.
We further give detailed advice on stabilizing RUF implementation and safe RUF usage to minimize RUF impacts on the Rust ecosystem based on our RUF findings.

\subsection{RUF Dependency Detection and Compilation Failure Recovery}
\label{RUF Impacts Mitigation}

\noindent 
\textbf{Problems.}
RUF can propagate through transitive dependencies, and package developers are usually unaware of such propagation.
As a result, when RUF dependency exposes vulnerabilities or reliability issues, developers can not easily fix the problems as they are introduced by dependencies, and dependency source code can not be directly modified.
Compilation failure is an example that removed RUF can cause through transitive dependencies, where the compiler only exports an error message, and developers often have no idea what happens when the RUF is introduced by dependencies.
Therefore, it is important to detect RUF dependency in advance to reveal RUF impacts and try to recover them from compilation failure to mitigate RUF impacts as much as possible.

\vspace{+4pt}
\noindent 
\textbf{Design.}
To further mitigate RUF impacts on the Rust ecosystem, we design a new tool to detect RUF dependency in packages and try to recover them from RUF threats, including compilation failure, unstable usage, etc.
The main mitigation strategy is to find a compatible version of the Rust compiler, where all types of RUF used by the package can keep their best status.
This can be achieved only when we maintain RUF status in every version of the compiler (RUF lifetime).
Without RUF lifetime, developers don't know how to choose compiler versions to keep RUF in its best status.
Our design uses the EDG generator to determine RUF dependencies and use RUF status extraction data to locate the compatible compiler for packages using RUF.
Taking \cref{tab:rufrem_exam} for example, the example package is impacted by RUF A\&B\&C. After traversing all compiler versions, in compiler version 1.57.0, all RUF used are at their best status, so this compiler version is selected for the package.

\vspace{+2pt}
\noindent \textbf{Implementation.}
We develop \textit{RUF dependency detector} of Rust projects, which gives analyzed packages their RUF dependency information, advice, and compatible Rust compiler version.
It is represented as a sub-command of Rust official package manager \textit{Cargo} for compatibility and can analyze the Rust projects with given arguments and options passed to Cargo.
This ensures that our tool can extract the exact environment of the compilation process, including running operating system, architecture, enabled PF, and other compiler flags. Under the environment, our tool will then use the technique proposed in \cref{fig:impl_arch} to accurately find all enabled RUF to
1) warn developers if enabled RUF might get removed in newer Rust compilers,
2) switch the package to a compatible compiler if it suffers from compilation failure introduced by enabled RUF,
3) detect RUF with an abnormal lifetime which is more vulnerable and unstable for the projects, and
4) audit final dependencies to avoid RUF impacts.
Due to the time limit, our implementation only includes functionality 2), but others can be easily extended as our code base and database include necessary functionality and data.

\begin{table}[]
    \caption{\data{RUF impact mitigation results.} Applying our mitigation strategy, \data{90\%} of package versions can recover from compilation failure. }
    \label{tab:rufrem}
    
    \resizebox{0.7\columnwidth}{!}{%
        \begin{tabular}{lcc}
        \hline
        RUF Impacts     & Total  & Compilation Failure \\ \hline
        Before Mitigation & 259,540 & 70,913               \\
        After Mitigation  & 259,540 & 6,978                \\ \hline
        \end{tabular}
    }
\end{table}

\vspace{+2pt}
\noindent \textbf{Mitigation Results.}
We develop the \textit{RUF mitigation analyzer} of the Rust ecosystem, which scans the Rust ecosystem to reveal the mitigation success rate of our tool. The results are shown in \cref{tab:rufrem}. 
Originally, there are \data{259,540} package versions impacted by RUF, and at most \data{70,913} package versions suffer from compilation failure in the newest Rust compiler in theory. Applying our compilation failure mitigation design, over \data{90\% (63,935/70,913)} of package versions can recover from compilation failure.
Specifically, we look into the super-spreader \code{redox_syscall-0.1.57} to access our mitigation results. We find that \data{97\% (40516/41750)} of its dependents successfully apply our mitigation technique and recover from compilation failure, which accounts for \data{63\% (40516/63935)} of our successfully recovered package versions. This reveals the significant desire for careful super-spreaders package maintenance to avoid single-point failure.
The mitigation result points out the effectiveness of our mitigation technique and proves that it can contribute to the reliability and usability of the Rust ecosystem.
However, we must add that this is not done once and for all. The RUF may contain other potential bugs and are not supported in other Rust compiler versions. As a result, the ultimate solution to avoid RUF impacts is to stabilize RUF and the development standard of RUF.
Our tool cannot change the stabilization process and can only select compatible compilers to help developers mitigate RUF impacts as much as possible.
\begin{flushleft}
    \fbox{\begin{minipage}[l]{.462\textwidth}
    \textbf{Mitigation Results:} Our proposed RUF impact mitigation technique can recover up to \data{90\%} of package versions in the Rust ecosystem that suffer from compilation failure introduced by RUF impact. 
    \end{minipage}}
\end{flushleft}

\subsection{Suggestions on RUF Impacts Mitigation}
\noindent 
\textbf{Stabilizing RUF Implementation.}
For the Rust compiler, we suggest that the development process of the compiler should be systematically reconsidered.
The introduction of RUF can cause stability problems as it is now widely used and impacts a large part of the whole ecosystem, which makes the ecosystem less stable. What's worse, the RUF compatibility problems can make the whole package unusable for developers outside of the ecosystem, which is not friendly for maintainers and open source software community. 
First, for compatibility concerns, we suggest that useless RUF should not be removed. If the RUF is not proven to be vulnerable or unstable, it can still work for developers, but with a warning instead.
Second, super-spreaders that a large number of packages depend on should avoid using RUF, especially when it is enabled by default. This can greatly eliminate RUF impacts in the Rust ecosystem.
Third, The Rust compiler should provide more information about RUF enabled to developers to build more reliable software.

\vspace{+2pt}
\noindent 
\textbf{Safely Using RUF.}
Although the Rust ecosystem declares itself to be stable for developers, lots of packages will then face compilation failure caused by removed RUF.
To stabilize the Rust ecosystem, we recommend that developers (especially those who manage popular packages) backport their fixes to old versions. Following semantic versioning, dependent packages can automatically use fixed versions to avoid such failure. 
Moreover, to avoid RUF usage as much as possible, developers should use RUF only when inevitable. The RUF usage should be limited to a set of data structures and functions to be efficiently replaced in the future. 
What's more, developers should enable RUF only when it is needed. Developers can use RUF configuration predicates to declare environment and functionality requirements to enable RUF.
Last, we suggest that developers audit their RUF usage and dependencies to avoid using RUF as much as possible. Developers can use audit tools (e.g., our tool in \cref{RUF Impacts Mitigation}) to help review their own projects.

\section{Threat to Validity}
\label{sec:validity}

First, in the evaluation process of our EDG, we mentioned we remove local configurations to keep consistent with the ecosystem behavior. However, we find that no more than \data{1\%} of packages can't be easily processed in this way. And the configurations can't be successfully resolved by \textit{Cargo Tree}, so we removed these packages in the ground truth, which makes the data set slightly smaller than it should be.
Second, \data{92\%} of RUF configuration predicates defined by Rust packages in the ecosystem can be successfully recognized. Others are considered as not impacting the ecosystem through dependencies. This makes RUF impacts underestimated.
\section{Related Work}
\label{sec:related}


\noindent \textbf{Dependency Analysis}.
Package managers (PMs) made different decisions on dependency definition and resolution.
Pietro et al. \cite{dependency_resolution} systematically compared resolvers in various dimensions, including conflict solutions, range modifiers, etc.
Jens et al. \cite{dependency_version} summarized forms of dependency version classifications under different PMs.
Decan et al. \cite{sem_dep} and Zhang et al. \cite{sem_dep} focus on the usage and compatibility issues of semantic versioning. 
Decan et al. \cite{dependency_network_comparision, OSS_dependency, topology_dependency} also defined evolution metrics of comprehensive dimensions and systematically analyzed and compared ecosystem dependency graph evolution from different PMs by empirical study.
Zimmermann et al. \cite{small_world} revealed security risks in the NPM ecosystem by analyzing dependencies, including fragile maintainers and unmaintained packages which are popular in the NPM.
Liu et al. \cite{npm_vul} further used NPM-specific principles to correctly resolve dependencies and revealed vulnerability impacts.

\vspace{+2pt}
\noindent \textbf{Ecosystem Analysis via Dependency}.
Wittern et al. \cite{wittern2016look} conducted the first large-scale analysis of the NPM ecosystem. By analyzing the topology of the popular JavaScript libraries, they found that NPM packages heavily rely on a core set of libraries.
Li et al. \cite{rust_yanked} found that almost half the packages adopt yanked releases, and these yanked releases propagated through dependency, causing unbuildable problems.
Jia et al. \cite{jia2021depowl} and Mukherjee et al. \cite{mukherjee2021fixing} focused on dependency incompatibilities issues within C/C++ and Python, and propose to detect and fix these issues to ensure the repeatability of the build.
Wang et al. \cite{java_dc, python_watchman, golang_hero, dc_java} studied the manifestation and repair patterns of dependency conflicts of three language ecosystems (i.e., Java, Python, and Golang), and developed tools for automatic detection, testing, and monitoring.
The above research work is conducted using data from a small proportion of the whole ecosystem due to efficiency problem.
We believe that our techniques can be applied to precise and efficient dependency resolution for further ecosystem-level study. Based on our generated ecosystem dependency graph, vulnerability propagation, dependency conflict detection, and other ecosystem-level research can be easily and comprehensively conducted.

\vspace{+2pt}
\noindent \textbf{Reliability Research on Compiler and Rust}.
Although developers rely on compilers to build reliable programs, compilers can also introduce extra vulnerabilities \cite{P4_compiler_vuln, compiler_sidechannel, compiler_vuln, compiler_software, compiler_correctness}.
Hohnka et al. \cite{compiler_vuln} pointed out that popular compilers can induce vulnerabilities like undefined behavior, side-channel attacks, persistent state violation, etc.
There is also reliability research on Rust to scan Rust bugs better and prevent developing unsafe codes.
Bae et al. \cite{rust_rudra} suggested three types of memory safety bug patterns in Rust and developed a tool to automatically detect bugs in the Rust ecosystem.
Astrauskas et al. \cite{rust_unsafe} empirically studied unsafe code usage in practice, concluding six purposes for using unsafe code and three Rust hypotheses that help make unsafe code safer.
Li et al. \cite{codelevel:mirchecker} presented a static analysis tool to detect runtime assertions failure and common memory-safety bugs, by analyzing Rust’s Mid-level Intermediate Representation (MIR).
Jiang et al. \cite{codelevel:rulf} proposed an approach based on an API dependency graph to automatically generate fuzz targets for fuzzing Rust library and implement a tool that can efficiently generate fuzz targets on a given library API and integrated with AFL++ for fuzzing.
These analysis techniques can also be applied to the Rust compiler to help stabilize the compiler codes and RUF implementations.
Our results show that, while Rust developers seek RUF for compiler functionality extensions, the reliability problems such as compilation failure and vulnerabilities can also be invisibly introduced to the Rust ecosystem on large scale. We hope that researchers and the Rust community can investigate further the compiler problems.


\section{Conclusion}
\label{sec:conclu}
In this paper, we conduct the first in-depth study to analyze RUF usage and its impacts on the Rust ecosystem.
We propose several novel techniques in the analysis, including {compilation data-flow interception}, {semantic
identification of RUF configuration}, {ecosystem dependency graph generator}, and {RUF impact mitigation}.
More specifically, we first extract the RUF definition from the compiler and usage from packages. Then we resolve all package dependencies for the entire ecosystem to quantify the RUF impacts on the whole ecosystem. We also remediate RUF impacts by finding out the best recovery point for impacted packages.
Our analysis covers the whole Rust ecosystem with \data{590K} package versions and \data{140M} transitive dependencies. 
Our study shows that \data{44\%} of package versions are affected by RUF, causing at most \data{12\%} of package versions to fail to compile. Moreover, using our RUF remediation technique, up to \data{90\%} package versions can be recovered from compilation failure caused by RUF impacts.
Our study discovers many useful findings and reveals the importance of stabilizing RUF for the security and reliability of the Rust ecosystem.

\begin{acks}
The authors would like to thank all reviewers sincerely for their valuable comments. This work is partially supported by the National Key R\&D Program of China (2022YFB3103900), by the
National Natural Science Foundation of China (Grant No. 62002317) and by the Hangzhou Leading Innovation and Entrepreneurship Team (TD2020003).
\end{acks}

\bibliographystyle{ACM-Reference-Format}
\bibliography{reference}

\end{document}